\documentstyle[aps,prl]{revtex}

\draft

\newcommand{\bm}[1]{{\mbox{\boldmath$#1$}}}

\begin{document}

\title{
   Mechanical effects in quantum dots in magnetic and electric fields}
\author{
   Lucjan Jacak$^a$,
   Jurij Krasnyj$^{bc}$, 
   Dorota Jacak$^d$,
   and Arkadiusz W\'ojs$^a$}
\address{
   $^a$Institute of Physics, Wroc{\l}aw University of Technology,
       Wybrze\.ze Wyspia\'nskiego 27, 50-370 Wroc{\l}aw, Poland \\
   $^b$Institute of Mathematics, University of Opole, Oleska 48, 
       Opole, Poland\\
   $^c$Institute of Physics, Odessa University, Odessa, Ukraine\\
   $^d$Institute of Mathematics, Wroc{\l}aw University of Technology,
   Wybrze\.ze Wyspia\'nskiego 27, 50-370 Wroc{\l}aw, Poland}
\maketitle
\begin{abstract}
   The mechanical effects in finite two-dimensional electron 
   systems (quantum dots or droplets) in a strong perpendicular 
   magnetic field are studied.
   It is shown that, due to asymmetry of the cyclotron dynamics, 
   an additional in-plane electric field causes a ground state 
   transition accompanied by a change in the average total angular 
   momentum of the system, unless the lateral confining potential
   is exactly parabolic.
   A precise mechanical experiment is proposed in which a 
   macroscopic angular momentum of a dense matrix of quantum 
   dots could be measured and used to detect and estimate 
   anharmonicity of the confinement.
\end{abstract}


\section{Introduction}

The manufacturing of semiconductor quantum dots with a controlled 
number of confined electrons allows the experimental study of finite 
quantum systems in the area not accessible in ordinary atomic physics 
(cf.\ \cite{jac} for review). 
Particularly interesting are the magnetic effects which in quantum 
dots are of the orders of magnitude greater than in atoms, as a
consequence of the magnetic length $l_{B}=\sqrt{\hbar c/eB}$ being 
comparable to the dot dimensions even at relatively small fields.
This is connected with the fact that the de Broglie wavelength of 
carriers in many semiconductors is significantly larger than the 
atomic dimensions. 
This property is also responsible for the quasi-two-dimensional 
(2D) character of the dynamics of quantum dots created in narrow 
semiconductor quantum wells \cite{tarucha}.
The lateral confining potential for small 2D dots is generally 
nonsingular, as opposed to ordinary atoms or ionized donors or 
acceptors in semiconductor heterostructures, and it can be often 
well approximated by an isotropic 2D harmonic well. 
High accuracy of this approximation for quantum dots was confirmed 
by numerous experiments (far-infrared spectroscopy\cite{hei} 
indicating the validity of the generalized Kohn theorem\cite{kohn} 
and the measurements of the addition spectrum in a magnetic field
\cite{tarucha,ash}), as well as by theoretical calculations (exact 
diagonalization methods applied to realistic models of quantum dots
\cite{arek}).
Quantum dots and quantum dot systems are currently under intensive 
investigation because of their potential application in lasers
\cite{laser} and because they seem to be promising candidates for 
hardware elements for quantum information processing\cite{div,rossi}. 
The flexibility of the electronic structure of quantum dots which
can be controlled for example by external magnetic and electric 
fields is of major importance for all these applications.

In this paper we consider shifts of energy, momentum and angular 
momentum resulting from asymmetry of the cyclotron motion of electrons 
in 2D quantum dots in the presence of perpendicular magnetic and 
in-plane electric fields. 
The vanishing of the mechanical effects for the special case of a 
parabolic lateral confinement makes mechanical experiments a potential 
new method (in addition to the optical experiments based on the Kohn 
theorem) to study the anharmonicity of the confinement. 
Even though the predicted shifts of mechanical momenta due to the 
anharmonicity of the confinement are very small for a single quantum 
dot, they can be significantly enhanced by using a dense (possibly 3D) 
matrix of dots.
It seems plausible that the measurement of a resulting macroscopic 
mechanical quantities could be possible in an ultra-precise experiment.

\section{2D quantum dot in perpendicular magnetic 
         and in-plane electric fields}

Let us consider a single 2D quantum dot containing $N$ electrons 
confined by a lateral potential and subject to a perpendicular 
magnetic field and an additional static in-plane electric field. 
This system is described by the Schr\"odinger equation:
\begin{equation}
   \hat{H}\Psi_\lambda(\bm{r}_1,\bm{r}_2,\dots,\bm{r}_N)
   =
   {\cal E}_\lambda \Psi_\lambda(\bm{r}_1,\bm{r}_2,\dots,\bm{r}_N),
\end{equation}
where $\lambda$ comprises all relevant many-body quantum numbers.
The total Hamiltonian $\hat{H}=\sum_{i=1}^N\hat{H}_i+\hat{H}_{\rm int}$ 
contains the Coulomb interaction term
\begin{equation}
   \hat{H}_{\rm int} 
   = \sum_{i<j=1}^N {e^2\over\epsilon_0|\bm{r}_i-\bm{r}_j|},
\end{equation}
and the single-particle term
\begin{equation}
   \hat{H}_i = {1\over2m}\left(\bm{p}_{\bm{r}_i} 
             + {e\over c}\bm{A}_{\bm{r}_i}\right)^2
             + V(\bm{r}_i)
             + e\bm{E}\bm{r}_i 
             + g\mu_B\bm{\sigma}_i\bm{B}
\label{eq_ham}
\end{equation}
whose eigenstates and energies are denoted by $\psi$ and $\varepsilon$,
\begin{equation}
   \hat{H}_i \psi(\bm{r}_i) = \varepsilon \psi(\bm{r}_i).
\label{eq_eigen1}
\end{equation}
In the above, $-e$ and $m$ are the electron charge and effective 
mass, $\bm{r}=(x,y,0)$ and $\bm{p}_\bm{r}=-i\hbar\,\partial/\partial
\bm{r}$ are the position and momentum operators, $V$ is the lateral 
confining potential, $\bm{B}=(0,0,B)$ and $\bm{A}_\bm{r}={1\over2}
\bm{B}\times\bm{r}$ are the external magnetic and its vector potential 
in the symmetric gauge, $\bm{E}=(E_x,E_y,0)$ is the external electric 
field, $\epsilon_0$ is the dielectric constant, and $\mu_B$ and $g$ 
are Bohr magneton and the gyromagnetic factor. 
We included the Pauli term even though for some materials the Zeeman 
splitting is very small compared to the typical orbital excitation 
energy of a few meV (e.g., for the bulk GaAs at small $B$ the Zeeman 
splitting is only about $0.03$~meV/T). 
This term has no influence on our results.

If the lateral confining potential is a 2D harmonic well,
\begin{equation}
   V(\bm{r}_i) = {1\over2}m\omega_0^2r_i^2,
\end{equation}
with a (arbitrary) characteristic frequency $\omega_0$, it is possible 
to remove formally the in-plane electric field from $\hat{H}$ by an 
appropriate coordinate shift and a simultaneous gauge transformation.
Let us first demonstrate it for the single-particle term, which can
be rewritten as
\begin{equation}
   \hat{H}_i
   = {1\over2m}
     \left(\bm{p}_{\bm{r}_i}+{e\over c}\bm{A}_{\bm{r}_i}\right)^2
   + {1\over2}m\omega_0^2(\bm{r}_i+\bm{r}_0)^2
   - {1\over2}m\omega_0^2r_0^2
   + g\mu_B{\bm\sigma}_i\bm{B},
\end{equation}
where the ``position shift'' is
\begin{equation}
   \bm{r}_0 = {e\bm{E}\over m\omega_0^2}.
\label{eq_shift}
\end{equation}
Now we formally change the coordinate, $\bm{r}_i+\bm{r}_0=\bm{x}_i$ 
(note that $\bm{p}_\bm{r}=\bm{p}_\bm{x}$) to obtain
\begin{equation}
   \hat{H}_i 
   = {1\over2m}
     \left(
        \bm{p}_{\bm{x}_i} + {e\over c}\bm{A}_{\bm{x}_i-\bm{r}_0}
     \right)^2
   + {1\over2}m\omega_0^2x_i^2
   - {1\over2}m\omega_0^2r_0^2
   + g\mu_B{\bm\sigma}_i\bm{B}
\end{equation}
and the eigenequation [cf.\ Eq.~(\ref{eq_eigen1})] 
\begin{equation}
   \hat{H}_i\psi(\bm{x}_i-\bm{r}_0) 
   = \varepsilon\psi(\bm{x}_i-\bm{r}_0).
\label{eq_eigen2}
\end{equation}
We can now use the linearity of the vector potential, 
\begin{equation}
   \bm{A}_{\bm{x}_i-\bm{r}_0}
   = \bm{A}_{\bm{x}_i} - {1\over2}\bm{B}\times\bm{r}_0,
\end{equation}
and the gauge invariance principle to write the gauge 
transformed Hamiltonian,
\begin{equation}
   \hat{H}'_i 
   = {1\over2m}
     \left(\bm{p}_{\bm{x}_i}+{e\over c}\bm{A}_{\bm{x}_i}\right)^2
   + {1\over2}m\omega_0^2x_i^2 
   - {1\over2}m\omega_0^2r_0^2
   + g\mu_B{\bm\sigma}_i\bm{B}.
\end{equation}
The eigenfunctions of $\hat{H}'_i$ satisfy
\begin{equation}
   \hat{H}'_i\psi'(\bm{x}_i) = \varepsilon\psi'(\bm{x}_i),
\end{equation}
with the same eigenenergies $\varepsilon$ as in Eqs.~(\ref{eq_eigen1}) 
and (\ref{eq_eigen2}), and $\psi'$ different from $\psi$ only by a phase 
factor,
\begin{equation}
   \psi(\bm{r}_i)
   = \exp\left[
     {ie\over2\hbar c}(\bm{B}\times\bm{r}_0)\cdot\bm{x}_i
     \right] \psi'(\bm{x}_i).
\label{eq_wave}
\end{equation}
Finally, we note that $\hat{H}'_i$ is different from $\hat{H}_i^o$,
the (axially symmetric) single-particle Hamiltonian $\hat{H}_i$ of 
Eq.~(\ref{eq_ham}) in the absence of an electric field, only by a 
constant,
\begin{equation}
   \hat{H}'_i = \hat{H}_i^o - {1\over2}m\omega_0^2r_0^2,
\end{equation}
and therefore the eigenfunctions $\psi'$ of the gauge transformed 
Hamiltonian $\hat{H}'_i$ satisfy
\begin{equation}
   \hat{H}_i^o \psi'(\bm{r}_i) = \varepsilon^o \psi'(\bm{r}_i)
\end{equation}
with the eigenenergies $\varepsilon$ given by
\begin{equation}
   \varepsilon = \varepsilon^o - {1\over2}m\omega_0^2r_0^2
               = \varepsilon^o - \Delta\varepsilon.
\label{eq_energy}
\end{equation}
The pair of Eqs.~(\ref{eq_wave}) and (\ref{eq_energy}) show that
the effect of an electric field $\bm{E}$ on the single-particle 
spectrum of a parabolic dot consists merely of a rigid displacement 
of the wavefunctions by $\bm{r}_0\propto\bm{E}$, a phase factor, 
and a constant shift of energies by $\Delta\varepsilon\propto E^2$.
In particular, despite breaking of the rotational symmetry by the
electric field, the wavefunctions remain rotationally symmetric
(although the axis of symmetry is displaced from the center of
the confining potential $V$).
Similar conclusions remain valid for the many-electron system since 
the interaction term $\hat{H}_{\rm int}$ of the many-body Hamiltonian 
$\hat{H}$ is translationally invariant,
\begin{equation}
     \sum_{i<j=1}^N{e^2\over\epsilon_0|\bm{r}_i-\bm{r}_j|}
   = \sum_{i<j=1}^N{e^2\over\epsilon_0|\bm{x}_i-\bm{x}_j|}.
\end{equation}
Therefore, as a consequence of the gauge invariance and the harmonic 
form of confinement $V$, the wavefunctions $\Psi_\lambda$ and energies 
${\cal E}_\lambda$ of the interacting many-electron system are:
\begin{equation}
   \Psi_\lambda(\bm{r}_1,\bm{r}_2,\dots,\bm{r}_N)
   = \exp\left[
     {ie\over2\hbar c}(\bm{B}\times\bm{r}_0)\cdot\sum_{i=1}^N\bm{x}_i
     \right] \Psi'_\lambda(\bm{x}_1,\bm{x}_2,\dots,\bm{x}_N)
\label{eq_gi_A}
\end{equation}
and
\begin{equation}
   {\cal E}_\lambda = {\cal E}'_\lambda - {1\over2}Nm\omega_0^2r_0^2,
\label{eq_gi_B}
\end{equation}
where $\Psi'_\lambda$ and ${\cal E}'_\lambda$ describe the system at 
$\bm{E}=0$,
\begin{equation}
   \hat{H}^o \Psi'_\lambda(\bm{r}_1,\bm{r}_2,\dots,\bm{r}_N)=
   {\cal E}'_\lambda \Psi'_\lambda(\bm{r}_1,\bm{r}_2,\dots,\bm{r}_N).
\label{eq_gi_C}
\end{equation}
Let us stress that the above-discussed symmetry of a multi-electron 
planar quantum dot with parabolic confinement is independent of the 
symmetry responsible for the Kohn theorem and connected with the 
separation of the center of mass and relative dynamics.

\section{Effects of an in-plane electric field 
         in parabolic quantum dots}

The transformation presented above allows the calculation of various 
effects caused by an in-plane electric field in the presence of 
a magnetic field oriented perpendicularly to the quantum dot.
For example, using Eqs.~(\ref{eq_gi_A})--(\ref{eq_gi_C}) one can 
calculate the change of the average energy $\left<{\cal E}\right>$ 
at a temperature $T$ by applying the canonical Gibbs distribution,
\begin{equation}
   \left<{\cal E}\right> 
   = {1\over{\cal Z}}{\rm Tr}\,[\hat{H}e^{-\beta\hat{H}}] 
   = {1\over{\cal Z}}\sum_\lambda{\cal E}_\lambda
     e^{-\beta{\cal E}_\lambda}
   = \left<{\cal E}'\right> 
   - {Ne^2E^2\over2m\omega_0^2},
\end{equation}
where
\begin{equation}
   \left<{\cal E}'\right> 
   = {1\over{\cal Z}'}\sum_\lambda{\cal E}'_\lambda
     e^{-\beta{\cal E}'_\lambda}
\end{equation}
describes the dot at $\bm{E}=0$.
Here, ${\cal Z}={\rm Tr}\,e^{-\beta\hat{H}}$ is the statistical function
and $\beta=(k_BT)^{-1}$.
The polarizability $\alpha$ is 
\begin{equation}
   \alpha = -{1\over E}{\partial\left<{\cal E}\right>\over\partial E}
          = {Ne^2\over m\omega_0^2}.
\end{equation}
For the matrix of $n$ quantum dots per unit area, the correction to 
the electric susceptibility is
\begin{equation}
   \Delta\epsilon={nNe^2\over m\omega_0^2}.
\end{equation}
Note that $\Delta\epsilon$ depends on neither $T$ nor $B$.
One can also calculate the expectation value of the generalized 
momentum,
\begin{eqnarray}
   \left<\bm{p}\right>_\lambda
  &=&\int
     \Psi^*_\lambda(\bm{r}_1,\bm{r}_2,\dots,\bm{r}_N)
     \left[\sum_{i=1}^N \bm{p}_{\bm{r}_i}\right]
     \Psi_\lambda(\bm{r}_1,\bm{r}_2,\dots,\bm{r}_N)\,
     d\bm{r}_1 d\bm{r}_2 \dots d\bm{r}_N
\nonumber\\
  &=&\int
     \Psi'^*_\lambda(\bm{x}_1,\bm{x}_2,\dots,\bm{x}_N)
     \left[\sum_{i=1}^N \bm{p}_{\bm{x}_i}\right]
     \Psi'_\lambda(\bm{x}_1,\bm{x}_2,\dots,\bm{x}_N)\,
     d\bm{x}_1 d\bm{x}_2 \dots d\bm{x}_N
   + {Ne\over2c}\bm{B}\times\bm{r}_0 
\nonumber\\
  &=&\left<\bm{p}'\right>_\lambda
   + {Ne\over2c}\bm{B}\times\bm{r}_0 
   = {Ne^2\over2cm\omega_0^2}\bm{B}\times\bm{E}.
\end{eqnarray}
Since expectation value is independent of $\lambda$, the statistical 
average at a temperature $T$ is the same, $\left<\bm{p}\right>=
\left<\bm{p}\right>_\lambda$.
The expectation value of the velocity operator\cite{lan} 
(gauge invariant kinetic quantity) vanishes,
\begin{eqnarray}
   \left<\bm{v}\right>_\lambda
 &=&\int\Psi^*_\lambda(\bm{r}_1,\bm{r}_2,\dots,\bm{r}_N)
   \left[
      \sum_{i=1}^N{1\over m}
      \left(
         \bm{p}_{\bm{r}_i}+{e\over c}\bm{A}_{\bm{r}_i}
      \right)
   \right]
   \Psi_\lambda(\bm{r}_1,\bm{r}_2,\dots,\bm{r}_N)\,
   d\bm{r}_1 d\bm{r}_2 \dots d\bm{r}_N
\nonumber\\
 &=&\int\Psi'^*_\lambda(\bm{x}_1,\bm{x}_2,\dots,\bm{x}_N)
   \left[
      \sum_{i=1}^N{1\over m}
      \left(
         \bm{p}_{\bm{x}_i}+{e\over c}\bm{A}_{\bm{x}_i}
      \right)
   \right]
   \Psi'_\lambda(\bm{x}_1,\bm{x}_2,\dots,\bm{x}_N)\,
   d\bm{x}_1 d\bm{x}_2 \dots d\bm{x}_N
\nonumber\\
 &=&\left<\bm{v}'\right>_\lambda=0.
\end{eqnarray}
As for the momentum, $\left<\bm{v}\right>=\left<\bm{v}\right>_\lambda$.
For the generalized angular momentum (conserved only in the absence of 
the in-plane electric field) we obtain
\begin{eqnarray}
   \left<\bm{L}\right>_\lambda
   &=&
   \int\Psi^*_\lambda(\bm{r}_1,\bm{r}_2,\dots,\bm{r}_N)
   \left[
      \sum_{i=1}^N
         \bm{r}_i\times\bm{p}_{\bm{r}_i}
   \right]
   \Psi_\lambda(\bm{r}_1,\bm{r}_2,\dots,\bm{r}_N)\,
   d\bm{r}_1 d\bm{r}_2 \dots d\bm{r}_N
\nonumber\\
   &=&
   \int\Psi'^*_\lambda(\bm{x}_1,\bm{x}_2,\dots,\bm{x}_N)
   \left[
      \sum_{i=1}^N
         \bm{x}_i\times\bm{p}_{\bm{x}_i}
   \right]
   \Psi'_\lambda(\bm{x}_1,\bm{x}_2,\dots,\bm{x}_N)\,
   d\bm{x}_1 d\bm{x}_2 \dots d\bm{x}_N
   -{Ne\over2c}\bm{r}_0\times(\bm{B}\times\bm{r}_0)
\nonumber\\
&=& 
   \left<\bm{L}'\right>_\lambda
   -{Ne^3\over2cm^2\omega_0^4}\bm{E}\times(\bm{B}\times\bm{E})
   =
   \left<\bm{L}'\right>_\lambda
   -{Ne^3E^2\over2cm^2\omega_0^4}\bm{B},
\label{eq_L}
\end{eqnarray}
and, for the change of the thermodynamic average, 
$\Delta\left<\bm{L}'\right>=\Delta\left<\bm{L}'\right>_\lambda$.
The kinetic (gauge invariant) angular momentum $\bm{M}=\bm{r}\times 
m\bm{v}$ does not change in the electric field,
\begin{eqnarray}
   \left<\bm{M}\right>_\lambda
   &=&
   \int
   \Psi^*_\lambda(\bm{r}_1,\bm{r}_2,\dots,\bm{r}_N)
   \left[\sum_{i=1}^N\bm{r}_i\times
      (\bm{p}_{\bm{r}_i}+{e\over c}\bm{A}_{\bm{r}_i})\right]
   \Psi_\lambda(\bm{r}_1,\bm{r}_2,\dots,\bm{r}_N)\,
   d\bm{r}_1 d\bm{r}_2 \dots d\bm{r}_N
\nonumber\\
   &=&
   \int
   \Psi'^*_\lambda(\bm{x}_1,\bm{x}_2,\dots,\bm{x}_N)
   \left[\sum_{i=1}^N\bm{x}_i\times
      (\bm{p}_{\bm{x}_i}+{e\over c}\bm{A}_{\bm{x}_i})\right]
   \Psi'_\lambda(\bm{x}_1,\bm{x}_2,\dots,\bm{x}_N)\,
   d\bm{x}_1 d\bm{x}_2 \dots d\bm{x}_N
   = \left<\bm{M}'\right>_\lambda
\end{eqnarray}
Note also that the electric field does not affect magnetization,
\begin{equation}
   {\partial{\cal E}_\lambda\over\partial B}
   = 
   {\partial{\cal E}'_\lambda\over\partial B}.
\end{equation}

\section{Anharmonic Effects}

The simple dependence of many-electron wavefunctions and energies on 
the electric field given by Eqs.~(\ref{eq_wave}) and (\ref{eq_energy}) 
depended critically on the harmonic form of the confinement $V$.
This simple dependence resulted in the insensitivity of a number of 
measurable quantities to the electric field, among them the kinetic 
angular momentum $\bm{M}$.
In realistic quantum dots, whose confinement is not exactly harmonic,
the electric field causes more than a rigid displacement and a phase 
change of the single- and many-particle wavefunctions.
The harmonic case is the only case in which the combination of the 
rotationally symmetric confining potential and the potential of the 
uniform in-plane electric field remains rotationally symmetric.
In all other cases, the shape of the joint single-particle potential 
depends on $\bm{E}$, and so do the single- and many-particle charge 
density profiles and a number of (in principle) measurable quantities, 
such as $\bm{M}$.
One example is that of larger dots whose confinement can be usually 
well approximated by a rotationally symmetric hard-wall potential.
In such systems, an electric field creates an asymmetric potential 
minimum within a dot, and the electrons are confined to a smaller area.
Another example is that of very shallow dots in which a strong electric 
field can even cause unbinding of electrons.
All these effects are well-known in the context of 1D confinement 
of electrons in quantum wells and heterojunctions.

If indeed experimentally measurable, the dependence of $M$ on $E$ 
could be a mechanical probe of the anharmonicity of the confinement.
The major difficulty could be a small magnitude of the changes of 
$M$ expected for quantum dots, which however can be increased by 
many orders of magnitude in a dense (3D, i.e.\ multi-layer) matrix 
of dots.
The specific values of the total $M$ of $n$ dots in a unit area 
depend on many factors such as $n$, $N$, $B$, or $V$, but in the 
simplest case of a very large $B$, in which so-called fractional 
quantum Hall droplets form in larger quantum dots, $M$ is of the 
order of $nN(N-1)\hbar$.
To estimate the order of magnitude of the change of $M$ one can 
use the expression (\ref{eq_L}) for the change of $L$.
Using the following parameters for a dense matrix of GaAs dots:
$m=0.067$, $n=10^{10}$~cm$^{-2}$, $\hbar\omega_0=3.3$~meV, $N=10$, 
$E=10^9$~V/cm, $B=10$~T, and $S=1$~cm$^2$, we obtain 
$\Delta\left<L\right>=10^{-5}$~g\,cm$^2$/s.
Note that the factor $\omega_0^{-4}$ in Eq.~(\ref{eq_L}) strongly
favors shallower confinement, typical for dots defined electrostatically
\cite{ash,in} by means of a pattern of electrodes grown over a 2D 
heterostructure.
Whether detection angular momentum as small as estimated above is 
possible or not, we find the idea of the microscopic motion of a great 
number of electrons causing a macroscopic rotation of a sample quite 
intriguing.
The most sensitive measurement would probably involve the resonance 
between the vibrations of a dot matrix suspended in the form of 
a torsion pendulum and the oscillations of an electric field. 

\section{Conclusion}

We have studied the effect of the in-plane electric field $\bm{E}$ 
on the wavefunctions and energies of many-electron systems confined 
in quasi-two-dimensional quantum dots in a perpendicular magnetic 
field.
We have shown that for the special case of the harmonic lateral 
confining potential the effect of the electric field is a mere 
displacement of the many-electron wavefunction in the direction 
of the field and a change of phase.
In consequence, a number of measurable quantities, such as the 
kinetic angular momentum $\bm{M}$ remain unchanged in the electric
field.
Since the lack of dependence of $\bm{M}$ on $\bm{E}$ is a unique 
property of the harmonic confinement, the change of angular momentum 
under the variation of $\bm{E}$ is a measure of the actual 
anharmonicity of this confinement.
An experiment in which the rotation of a dense quantum dot matrix
under oscillation of an electric field occurs is proposed.

\section*{Acknowledgment}
This work was supported by KBN Project No: 2 PO3 B05518.

\end{document}